\documentclass[a4paper]{article}

\usepackage{INTERSPEECH2022}
\usepackage{booktabs}
\usepackage[table,x11names]{xcolor}
\usepackage[capitalize,noabbrev]{cleveref}

\newcolumntype{H}{>{\setbox0=\hbox\bgroup}c<{\egroup}@{}}

\title{FFC-SE: Fast Fourier Convolution for Speech Enhancement}
\name{ Ivan Shchekotov$^{*12}$, Pavel Andreev$^{*123}$, Oleg Ivanov$^1$, Aibek Alanov$^{124}$, Dmitry Vetrov$^{24}$}
\address{
  $^*$ Equal contribution \\
  $^1$ Samsung AI Center, Moscow\\
  $^2$ Higher School of Economics, Moscow \\
  $^3$ Skolkovo Institute of Science and Technology, Moscow \\
  $^4$ Artificial Intelligence Research Institute, Moscow}
\email{i.shchekotov@partner.samsung.com, p.andreev@samsung.com}

\begin{document}

\maketitle
\begin{abstract}
  Fast Fourier convolution (FFC) is the recently
proposed neural operator showing promising performance in several computer vision problems. The FFC operator allows employing large receptive field operations within early layers of the neural network. It was shown to be especially helpful for inpainting of periodic structures which are common in audio processing. In this work, we design neural network architectures which adapt FFC for speech enhancement. 
We hypothesize that a large receptive field allows these networks to produce more coherent phases than vanilla convolutional models, and validate this hypothesis experimentally. 
We found that neural networks based on Fast Fourier convolution outperform analogous convolutional models and show better or comparable results with other speech enhancement baselines. 
\end{abstract}

\section{Introduction}

Speech enhancement is of major interest in the audio processing community, as it has a fundamental importance in telecommunication. 
There are a lot of solutions for this problem in traditional signal processing, but each such solution relies on some assumptions on the underlying noise model.
Due to the recent advances in deep learning, data-driven approaches have dominated the area of modern speech enhancement. 

One popular line of deep learning techniques tackling speech enhancement is based on the time domain signal retrieval. These approaches often utilize a convolutional encoder-decoder (CED) structure. For example, \cite{tagliasacchi2020seanet} and \cite{pascual2017segan} follow an adversarial training pipeline and use a CED network as a generator employing a fully-convolutional discriminator for training. Some of these approaches additionally use neural modules that can capture long-range temporal sequence information, such as long short-term memory cells \cite{defossez2020real} and transformers~\cite{kim21h_interspeech}. However, since these techniques directly map a noisy waveform to the clean one, they typically leave aside any information about signal spectrum, causing potential inefficiencies.
One recent attempt to explicitly take into account spectral information during generation is \cite{hifiplus}. The authors propose a universal model for vocoding, speech enhancement, and bandwidth extension that takes as inputs both waveform and magnitude spectrogram and achieves state-of-art results. We show that the quality of speech enhancement can be further improved by our models.

Another line of research is built upon the estimation of short-time Fourier transform (STFT) spectrogram. 
Approaches of these lines aim to predict STFT coefficients of clean signal directly \cite{hao2020fullsubnet}~or correct spectrum of the noisy signal by estimating various masks for modification of magnitudes or both magnitudes and phases~\cite{choi2018phase, williamson2015complex,sun2017multiple}. 
For instance, MetricGAN~\cite{fu2019metricgan} and MetricGAN+~\cite{fu2021metricgan+} papers use Bidirectional LSTM to predict binary masks for magnitude spectrogram optimizing common speech quality objective metrics directly, and report state-of-the-art results for these metrics.
The direct estimation of spectrogram phases is challenging. Different tricks are proposed to simplify this task. 
These techniques include 
decoupling magnitude and phase estimation~\cite{kong2021decoupling}, and the usage of separate vocoder networks for waveform synthesis~\cite{liu2021voicefixer}. 
However, these methods tend to use large neural networks, requiring substantial computational resources.
We found that one of the limiting factors for phase prediction is local receptive field of these networks, preventing effective usage of models parameters.
We observed that phase estimation can be significantly facilitated by non-local neural operators, leading to much smaller model sizes while achieving better quality.

\begin{figure}[!h]
  \centering
  \includegraphics[width=\linewidth]{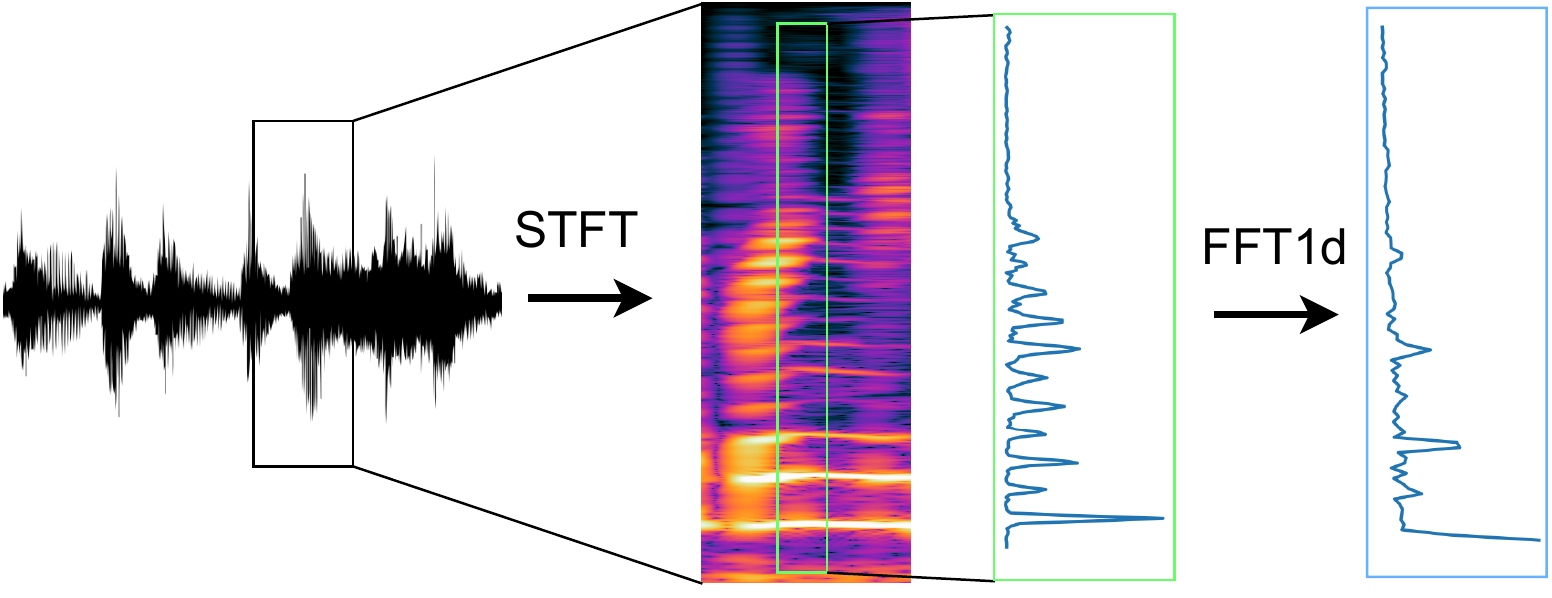}
  \caption{Harmonics of short-time Fourier transform spectrogram constitute periodic structures which can be naturally processed in Fourier domain by global branch of fast Fourier convolution.}
  \label{fig:ill}
   \vspace{-0.4cm}
\end{figure}

We propose new neural architectures based on fast Fourier convolution (FFC) operator \cite{chi2020fast} which we adapt for speech enhancement problems.
The FFC layers were originally proposed for computer vision tasks as a non-local operator replacing vanilla convolutional layers within existing neural networks.
Fast Fourier convolution has the global receptive field and was shown to be helpful for the restoration of periodic backgrounds in inpainting problems \cite{suvorov2022resolution}.
These properties of FFC are especially helpful for the spectrogram prediction.
Indeed, the harmonics of magnitude spectrogram are known to form periodic structures which can be naturally handled by fast Fourier convolution (see \Cref{fig:ill}).
Besides, we experimentally observe that a large receptive field is useful for producing coherent phases.
Based on these insights, we design new neural architectures for direct spectrogram estimation in speech enhancement problems. 
The proposed models achieve state-of-art performance on VoiceBank-DEMAND \cite{valentini2017noisy} and Deep Noise Suppression \cite{dubey2022icassp} datasets with much fewer parameters than the baselines.
The implementation will become publicly available.

\section{Proposed method}
We consider the standard single-channel speech denoising problem. In other words, our goal is to learn a mapping from noisy waveform $y = x + n$ with additive noise $n$ to the clean one $x$. We tackle this problem by neural architectures equipped with a non-local neural operator named fast Fourier convolution \cite{chi2020fast}. We adopt this operator for spectrogram processing and propose two neural architectures which use this operator as a basic block.

\subsection{Fast fourier convolution}
Fast Fourier convolution (FFC) \cite{chi2020fast} is a neural operator that allows performing non-local reasoning and generation within a neural network. FFC uses channel-wise fast Fourier transform \cite{nussbaumer1981fast} followed by a point-wise convolution and inverse Fourier transform, thus it globally affects input tensor across
dimensions involved in Fourier transform. 
FFC splits channels into local and global branches. 
The local branch uses conventional convolutions for local updates of feature maps.
Global branch performs Fourier transform of the feature map and updates it in spectral domain affecting global context.

\begin{figure}[!h]
  \centering
  \includegraphics[width=0.8\linewidth]{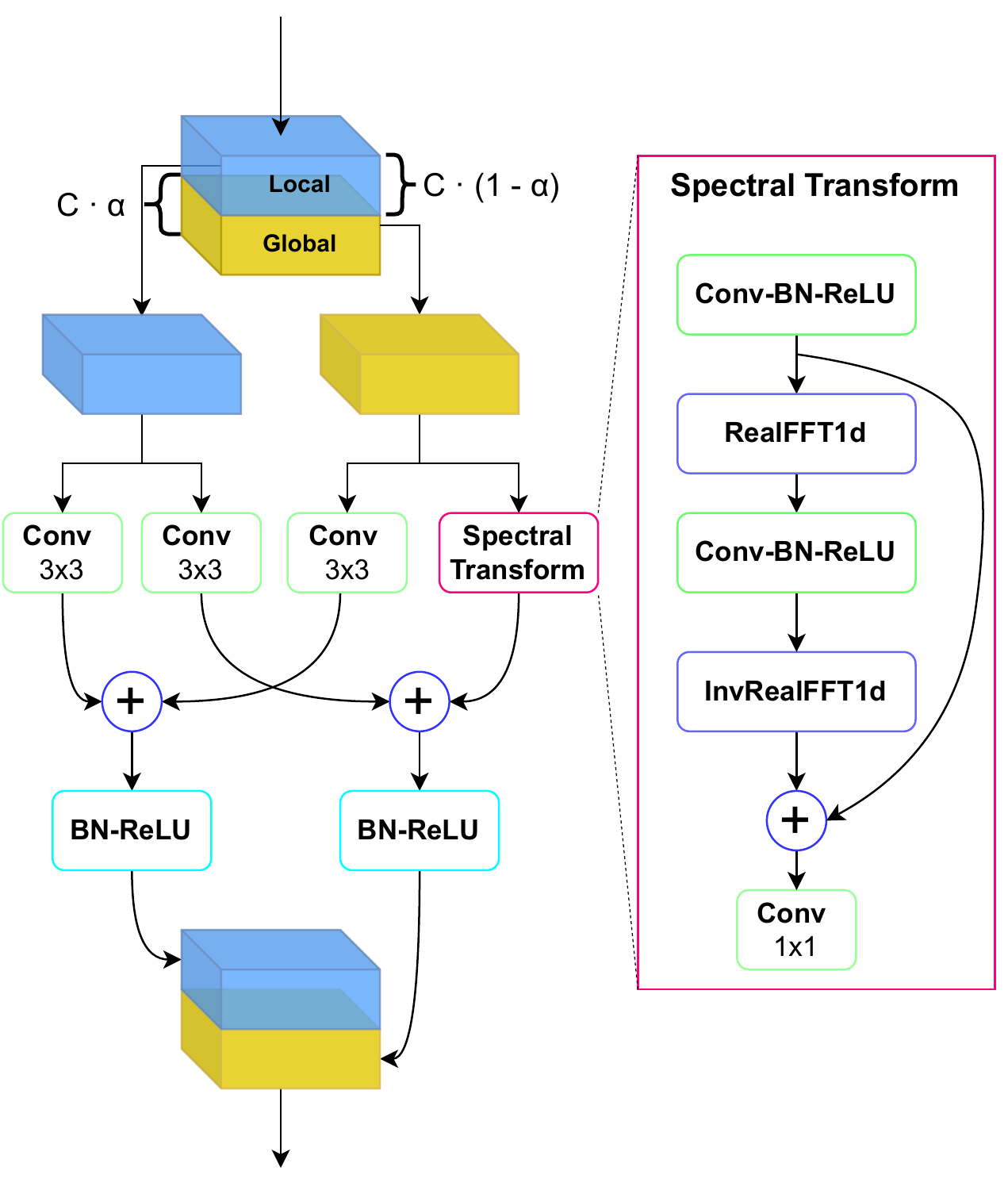}
  \caption{Fast Fourier Convolution neural module for speech enhancement. Parameter $\alpha \in [0, 1]$ controls the ratio of channels used in the global branch of the module.}
  \label{fig:ffc}
  	 \vspace{-0.3cm}
\end{figure}

In this work, we perform Fourier transform across frequency dimensions of feature maps (corresponding to STFT spectrograms) only  (see \Cref{fig:ill}), whereas in computer vision the Fourier transform is applied across both image dimensions \cite{chi2020fast, suvorov2022resolution}. Specifically, we implement the global branch of the FFC layer in three steps:
\begin{enumerate}
    \item Apply real unidimensional fast Fourier transform across frequency dimension of the input feature map and concatenate real and imaginary parts of spectrum across channel dimension: 
    $$\mathbb{R}^{C \times F \times T} \overset{\mathrm{fft1d}}{\longrightarrow} \mathbb{C}^{C \times F / 2 \times T} \overset{\mathrm{concat}}{\longrightarrow} \mathbb{R}^{2C \times F / 2 \times T} .$$
    \item Apply convolutional block (with $1\times 1$ kernel) in the frequency domain:
    $$\mathbb{R}^{2C \times F / 2 \times T} \overset{\mathrm{conv-bn-relu}}{\longrightarrow} \mathbb{R}^{2C \times F / 2 \times T} .$$
    \item Apply inverse Fourier transfrom:
   $$ \mathbb{R}^{2C \times F / 2 \times T} \overset{\mathrm{concat}}{\longrightarrow} \mathbb{C}^{C \times F / 2 \times T}  \overset{\mathrm{ifft1d}}{\longrightarrow} \mathbb{R}^{C \times F \times T} .$$
\end{enumerate}
where $C$, $F$, $T$ are the number of channels, dimension corresponding to frequency and dimension corresponding to time, respectively.
Global and local branches interact with each other through summation of activations, as illustrated in \Cref{fig:ffc}. We use the same variation of FFC that was explored in \cite{suvorov2022resolution} for image inpainting, except we utilize unidimensional Fourier transform across the frequency dimension.

\begin{figure*}[!h]
  \centering
  \includegraphics[width=0.8\linewidth]{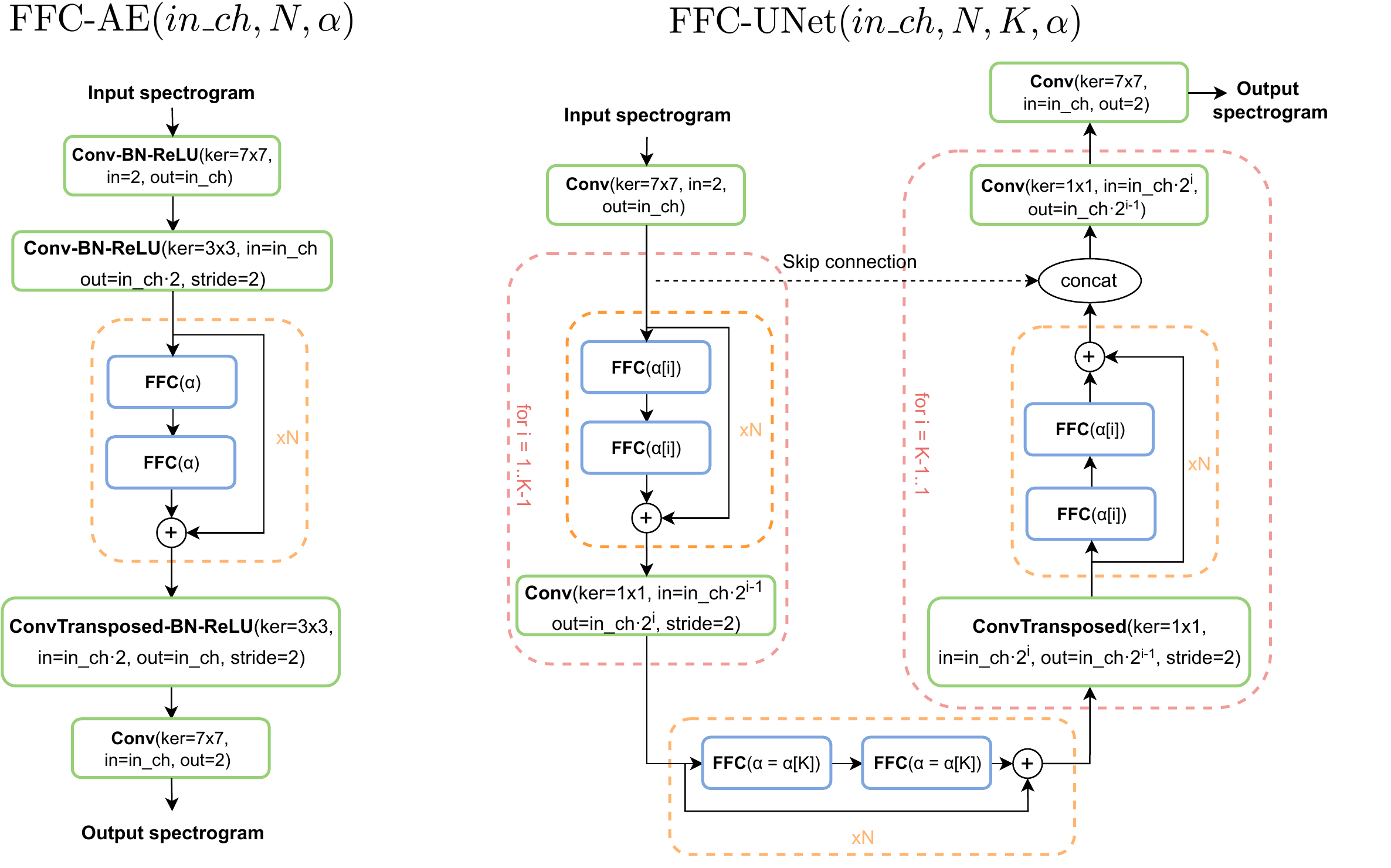}
  \caption{Proposed architectures for speech enhancement. \textit{Left:} fast Fourier convolutional autoencoder which adopts architecture introduced in \cite{suvorov2022resolution} for speech enhancement task. \textit{Right:} fast Fourier convolutional U-Net. Parameter in\_ch controls the overall width of the networks, N defines the number of FFC residual blocks, $K$ is the depth of the FFC-UNet architecture, $\alpha$ (real number $\in [0,1]$ in case of FFC-AE, K numbers $\in [0,1]$ in case of FFC-Unet) controls the proportion of channels going to the global branch.}
  \label{fig:ffc-arch}
   \vspace{-0.2cm}
\end{figure*}

\subsection{FFC-AE}
We implement two neural network architectures for speech enhancement. The first one (FFC-AE) is inspired by \cite{suvorov2022resolution}. This architecture consists of the convolutional encoder (strided convolution) which downsamples the input spectrogram across time and frequency dimensions by a factor of two. The encoder is followed by a series of residual blocks, each consisting of two sequential fast Fourier convolution modules. The output of residual blocks is then upsampled by transposed convolution and used to predict real and imaginary parts of the denoised spectrogram. The architecture is depicted on \Cref{fig:ffc-arch} (left). We call this model a fast Fourier convolutional autoencoder (FFC-AE).

Although bigger downsampling factors lead to a further reduction in the number of operations during inference, we found that it also leads to significant performance degradation, while factor 2 provides a good trade-off between performance and complexity. 
  \vspace{-0.2cm}
\subsection{FFC-UNet}
\label{subsec:ffc_unet}

The second architecture is inspired by the classic work \cite{ronneberger2015u}.
We incorporate FFC layers into U-Net architecture as shown in \Cref{fig:ffc-arch} (right). 
At each level of the U-Net structure, we utilize several residual FFC blocks with convolutional upsampling or downsampling. 
We find it beneficial to make the parameter $\alpha$ (ratio of channels going to a global branch of fast Fourier convolution) dependent on the U-Net level at which FFC is used.
Higher levels of U-Net structure work with higher resolutions of data at which periodic structures are present, while lower levels work at a coarse scale that lacks periodic structure. 
More generally, as noted in \cite{chi2020fast} the deeper layers of neural networks are mainly supposed to exploit local patterns, while the topmost layers highly demand contextual inference.
Thus, the global branch of FFC layers is less useful at the coarse scales and we decrease the parameter $\alpha$ starting from 0.75 at the topmost level to 0 at the bottom layer with step 0.25.  

\subsection{Training}

The predicted short-time Fourier transform spectrogram is converted into waveform by inverse short-time Fourier transform. 
We use the multi-discriminator adversarial training framework proposed in \cite{hifiplus} for time-domain models' training. 
It consists of three losses, namely LS-GAN loss $\mathcal{L}_{GAN}$ \cite{mao2017least}, feature matching loss $\mathcal{L}_{FM}$ \cite{larsen2016autoencoding, kumar2019melgan}, and mel-spectrogram loss $\mathcal{L}_{Mel}$ \cite{kong2020hifi}:
\begin{align}
    \mathcal{L}(\theta) &= \mathcal{L}_{GAN}(\theta) + \lambda_{fm}\mathcal{L}_{FM}(\theta) + \lambda_{mel}\mathcal{L}_{Mel}(\theta) \\
    \mathcal{L}(\varphi_i) &= \mathcal{L}_{GAN}(\varphi_i), \quad i = 1, \dots, k.
\end{align}
where $\mathcal{L}(\theta)$ denotes loss for generator with parameters $\theta$, $\mathcal{L}(\varphi_i)$ denotes loss for i-th discriminator with parameters $\varphi_i$ (all discriminators are identical). In all experiments we set $\lambda_{fm} = 2$, $\lambda_{mel} = 45$, $k=3$.

  \vspace{-0.1cm}
\section{Experiments and Results}
  \vspace{-0.05cm}
\begin{table*}[t]
    \centering
    \caption{Speech denoising results on Voicebank-DEMAND dataset. Best three results are highlighted in bold.}
    	 \vspace{-0.1cm}
	\label{tab:VB_main_res}
	\begin{tabular}{l c c c c c c c c | c H}
		\toprule
		Model & MOS & WV-MOS & SI-SDR & STOI & PESQ & CSIG & CBAK & COVL & \# Params (M) & \# GMAC on 16k \\
		\midrule
		\midrule
		Ground Truth & $4.46 \pm 0.06$ & 4.50 & - & 1.00 & 4.64 & 5.0 & 5.0 & 5.0 & - & -\\
		Input &  $3.44 \pm 0.06$& 2.99 & 8.4 & 0.79 & 1.97  & 3.34 & 2.82 & 2.74 & - & - \\
		\midrule
		MetricGAN+ \cite{fu2021metricgan+} & $3.82 \pm 0.06$  & 3.90 & 8.5 & 0.83 & \textbf{3.13} & 4.12 & 3.16 & 3.62 & 2.7 & - \\
		ResUNet-Decouple+ \cite{kong2021decoupling} & $3.94 \pm 0.04$ & 4.13 & \textbf{18.4} & 0.84 & 2.45 & 3.38 & 3.15 & 2.89 & 102.6 & - \\
		DEMUCS (non-caus.) \cite{defossez2020real} & $4.06 \pm 0.03$& \textbf{4.37} & \textbf{18.5} & \textbf{0.87} & \textbf{3.03} & \textbf{4.36} & \textbf{3.51} & \textbf{3.72} & 60.8 & - \\
				VoiceFixer \cite{liu2021voicefixer} & $4.10 \pm 0.03$ & 4.14 & -18.5 & 0.75 & 2.38 & 3.6 & 2.37 & 2.96 & 122.1 & 34.4 (x2) \\
			HiFi++ \cite{hifiplus} & $4.15 \pm 0.07$ & 4.27 & \textbf{18.4} & 0.86 & 2.76 & 4.09 & 3.35 & 3.43 & \textbf{1.7} & 1.5(x2) \\
		\midrule
		FFC-AE-V0 (ours) & $\mathbf{4.24 \pm 0.09}$ &  4.34 & 17.9 & 0.86 & 2.88 & 4.25 & 3.40 & 3.57 & \textbf{0.42} & 4.39 \\
		FFC-AE-V1 (ours) & $\mathbf{4.33 \pm 0.03}$  & \textbf{4.37} & 17.5 & \textbf{0.87} & 2.96 & \textbf{4.34} & \textbf{3.42} & \textbf{3.66}  & \textbf{1.7} & 16.33 \\
		FFC-UNet (ours) & $\mathbf{4.28\pm0.03}$ & \textbf{4.38} & 18.1 & \textbf{0.87} & \textbf{2.99} & \textbf{4.35} & \textbf{3.47} & \textbf{3.69} & 7.7 & 19.81 \\
		\midrule
		 FFC-AE-V1 (abl.) & $3.98 \pm 0.07$& 4.05 & 16.7 & 0.84 & 2.68 & 3.94 & 3.23 & 3.31 & 2.9 & 2.25 \\
		 vanilla UNet  & $4.10 \pm 0.07$ & 4.11 & 17.2 & 0.85 & 2.73 & 3.94 & 3.28 & 3.34 & 20.7 & 11.2(x2)  \\
		\bottomrule
	\end{tabular}
	 \vspace{-0.1cm}
\end{table*}
\begin{table*}[t]
    \centering
    \caption{Speech denoising results on DNS dataset. * indicates results on DNS-BLIND. Best three results are highlighted in bold.}
     \vspace{-0.1cm}
	\label{tab:DNS_main_res}
	\begin{tabular}{l  c c c c c c c c c c | c H}
		\toprule
		Model & MOS & MOS* & WV-MOS & WV-MOS* & SI-SDR & STOI & PESQ & CSIG & CBAK & COVL  & \# Params (M) & \# GMAC on 16k \\
		\midrule
		\midrule
		Ground Truth & $4.40 \pm 0.08$ & - &  3.845 & - & - & 1.00 & 4.64 & 5.0 & 5.0 & 5.0 & - & - \\
		Input & $2.75 \pm 0.07$ & $2.43 \pm 0.08$ & 1.195 & 0.80 & - &  0.69 & 1.49 & 2.59 & 2.32 & 1.99 & - & - \\
		\midrule

		DEMUCS  \cite{defossez2020real}& $3.52 \pm 0.15$ & $2.94 \pm 0.08$ & \textbf{3.32} & \textbf{2.83} & \textbf{15.56} & 0.82 & 2.20 & 3.44 & 3.21 & 2.81 & 33.5 & 7.84 \\
			HiFi++ \cite{hifiplus} & $3.54\pm 0.08$ & $2.75 \pm 0.06$ & 2.91 & 2.32 & 11.69 & 0.82 & 2.20 & 3.65 & 3.00 & 2.92 & \textbf{1.7} &  - \\
				ResUNet-Dec+ \cite{kong2021decoupling} & $3.63 \pm 0.04$ & $2.51 \pm 0.08$ &  2.94 & 1.86 & 14.78 & 0.81 & 2.09 & 2.82 & 3.06 & 2.43 & 102.6 & -   \\
		FullSubNet \cite{hao2021fullsubnet} & $3.73\pm0.02$ & $\mathbf{3.08 \pm 0.09}$ &  2.90 & 2.41 & \textbf{14.96} & 0.82 & 2.43 & 3.59 & \textbf{3.27} & 3.0 & 5.6 & - \\
        \midrule
        FFC-AE-V0 (ours) & $\mathbf{3.92 \pm 0.09}$ & $2.88 \pm 0.09$ & 3.20 & 2.58 & 12.86 & \textbf{0.83} & \textbf{2.44} &\textbf{ 3.84} & 3.17 & \textbf{ 3.15} & \textbf{0.42} & 4.39 \\
        FFC-AE-V1 (ours) & $\mathbf{4.02 \pm 0.05}$ & $\mathbf{3.10 \pm 0.07}$ & \textbf{3.33} & \textbf{2.76} & 14.12 & \textbf{0.85} & \textbf{2.61} & \textbf{3.98} & \textbf{3.31 }& \textbf{3.31} & \textbf{1.7} & 16.33 \\
		FFC-UNet (ours) & $\mathbf{4.00 \pm 0.06}$ & $\mathbf{3.11 \pm 0.08}$ &  \textbf{3.35} & \textbf{2.70} & \textbf{15.48} & \textbf{0.86} & \textbf{2.69} & \textbf{4.08} & \textbf{3.44} & \textbf{3.41} & 7.7 & 19.81 \\
		\bottomrule
	\end{tabular}
	 \vspace{-0.3cm}
\end{table*}

\subsection{Datasets}
We use two benchmarks for the evaluation of the effectiveness of the proposed speech denoising models. 

The first one is VoiceBank-DEMAND dataset \cite{valentini2017noisy} which is a standard benchmark for speech denoising systems.  
The train set consists of 28 speakers with 4 signal-to-noise ratios (SNR) (15, 10, 5, and 0
dB) and contains 11572 utterances. The test set (824 utterances) consists of 2 speakers unseen by the model during training with 4 SNR (17.5, 12.5, 7.5, and 2.5 dB). 

The second benchmark is the Deep Noise Suppression (DNS) challenge \cite{dubey2022icassp}. We synthesize 100 hours of training data using provided codes and default configuration. The only modification is that we do not utilize artificial reverberation during synthesis. The models are tested on two kinds of test sets. The first one (DNS-INDOMAIN) is a hold-out data randomly selected and excluded from synthesized 100 hours of training data. The second one (DNS-BLIND) is a standard blind test set from the DNS repository. This data is recorded in the presence of noise in real-world scenarios.
  \vspace{-0.05cm}
\subsection{Metrics}
  \vspace{-0.05cm}
\textbf{Objective metrics} We use conventional metrics WB-PESQ \cite{rix2001perceptual}, extended STOI \cite{jensen2016algorithm}, scale-invariant signal-to-distortion ratio (SI-SDR) \cite{le2019sdr}, COVL, CBAK, CSIG \cite{hu2007evaluation} for objective evaluation of samples in the concerned tasks. Metrics for all baselines and our models were calculated using the publicly available implementations and were not reused from original papers. In addition to conventional speech quality metrics, we considered absolute objective speech quality measure based on direct MOS score prediction by a fine-tuned wave2vec2.0 model (WV-MOS), which was found to have better system-level correlation with subjective quality measures than the other objective metrics \cite{hifiplus}. 

\noindent \textbf{Subjective metrics}  We use 5-scale MOS tests for subjective quality evaluation following procedure described in \cite{hifiplus}. 
All audio clips were normalized to prevent the influence of audio volume differences on the raters. 
The referees were restricted to be English speakers with proper listening equipment. 
  \vspace{-0.05cm}
\subsection{Experimental Setup}
  \vspace{-0.05cm}
In all our experiments, signals are transformed to the spectral domain using STFT with Hann window of size $1024$ and hop size $256$.
For FFC-AE model we set $\alpha = 0.75$, $N = 9$, $in\_ch = 32$ and $in\_ch = 64$ for V0 and V1 versions, respectively. For FFC-UNet $K = 4, N = 4$, $in\_ch = 32$ and $\alpha$ is gradually decreased with depth as described in Section \ref{subsec:ffc_unet}. Models are trained for 800 000 iterations with batch size being equal to 8. Adam optimizer is used with learning rate $0.0002$. ResUNet-Decouple+ \cite{kong2021decoupling} was trained with the same loss function as reported in the original paper, the number iterations was set 800 000 and learning rate to 0.0002.

\subsection{Experimental Results}
In addition to baselines from the literature \cite{fu2021metricgan+, kong2021decoupling, liu2021voicefixer, hifiplus, defossez2020real, hao2020fullsubnet}, we compare against fully convolutional U-Net model (vanilla U-Net) and a model which is the same as FFC-AE except for all Fourier units at the global branch are replaced with vanilla convolutions (FFC-AE (abl.)). These models follow exactly the same training setup as the proposed models for a clear illustration of the FFC importance.

\noindent \textbf{Phase estimation} We test the ability of the FFC-AE model to estimate phases given magnitude spectrograms on the LJ-Speech dataset \cite{ljspeech17} and compare against analogous architectures with vanilla convolutions.  The models were trained to predict phases (sine and cosine) and were provided with full magnitude spectrograms. The results are shown in \Cref{tab:phase}. FFC-AE significantly outperforms FFC-AE (abl.) and vanilla UNet models while having fewer parameters.

\begin{table}[!h]
  \vspace{-0.1cm}
    \centering
    \caption{Phase estimation on LJ-Speech dataset}
      \vspace{-0.1cm}
	\label{tab:phase}
	\begin{tabular}{l c c  | c H}
		\toprule
		Model & MOS &WV-MOS  & \# Params (M) & \# GMAC on 8k \\
		\midrule
		\midrule
		Ground Truth & 4.51 $\pm$ 0.05 & 4.23 & - & - \\
		\midrule
		FFC-AE-V0 (ours) & $\mathbf{4.47\pm 0.04}$  & \textbf{4.11} & 0.4 & - \\
		vanilla UNet & $4.31\pm 0.04$ & 3.97 & 20.7 & - \\
		FFC-AE-V0 (abl.) & $3.96 \pm 0.08$ & 3.81 & 0.7 & - \\
			\bottomrule
	\end{tabular}
	 \vspace{-0.1cm}
\end{table}

\noindent \textbf{Speech enhancement} We compare the quality of the proposed models with strong baselines on both benchmarks. On Voicebank-DEMAND, as it can be seen from  \Cref{tab:VB_main_res}, our models significantly outperform all the baselines by MOS and give competitive results on objective metrics. On DNS benchmark ( \Cref{tab:DNS_main_res}) our models have better quality than all the competitors considering DNS-INDOMAIN test set and perform competitively with FullSubNet \cite{hao2021fullsubnet} (one of the top-ranked models in DNS Challenge 2021) in terms of MOS on DNS-BLIND test set.

Noteworthy, our models performed better or comparably with the closest baselines FullSubNet and DEMUCS on DNS-BLIND set without employing dynamic data synthesis, reverberation simulation and augmentation techniques. Thus, the DEMUCS and FullSubNet models which employ these techniques were in an advantageous position from this point of view. We believe that the generalization of our models to the blind test set can be further improved, considering more advanced data generation pipelines.

  \vspace{-0.1cm}

\section{Conclusions}
 \vspace{-0.1cm}
In this paper, we adapted the fast Fourier convolution operator for speech enhancement problems.
We observe that neural architectures built upon fast Fourier convolution significantly outperform vanilla convolution-based architectures in terms of quality of speech enhancement, phase estimation, and parameter efficiency. 
In general, the proposed architectures deliver state-of-art results on speech denoising benchmarks, being significantly smaller than the baselines.
Future work should consider extending the results to real-time streaming scenarios.
Importantly, we believe that the success of fast Fourier convolution can be translated to other speech processing tasks, such as voice conversion and neural vocoding.
\section{Acknowledgements}
This work was supported by Samsung Research.

\newpage

\bibliographystyle{IEEEtran}

\bibliography{mybib}


\end{document}